# Thermodynamics of Phase Formation in the Quantum Critical Metal Sr$_3$Ru$_2$O$_7$

A.W. Rost[a,1], S.A. Grigera[a,b], J.A.N. Bruin[a], R.S. Perry[c], D. Tian[a], S. Raghu[d], S.A. Kivelson[e,1] & A.P. Mackenzie[a,1]

[a] SUPA, School of Physics and Astronomy, University of St Andrews, St Andrews KY16 9SS, United Kingdom

[b] Instituto de Física de Líquidos y Sistemas Biológicos, UNLP-CONICET, La Plata 1900, Argentina

[c] SUPA, School of Physics, University of Edinburgh, Mayfield Road, Edinburgh EH9 3JZ, United Kingdom

[d] Department of Physics and Astronomy, Rice University, Houston, Texas, 77005, USA

[e] Department of Physics, Stanford University, Stanford, California 94305, USA

[1] To whom correspondence should be addressed, to ar35@st-and.ac.uk, Kivelson@stanford.edu or apm9@st-and.ac.uk

**Abstract**

*The behaviour of matter near zero temperature continuous phase transitions, or 'quantum critical points' (QCPs) is a central topic of study in condensed matter physics. In fermionic systems, fundamental questions remain unanswered: the nature of the quantum critical regime is unclear because of the apparent breakdown of the concept of the quasiparticle, a cornerstone of existing theories of strongly interacting metals. Even less is known experimentally about the formation of ordered phases from such a quantum critical 'soup'. Here, we report a study of the specific heat across the phase diagram of the model system $Sr_3Ru_2O_7$, which features an anomalous phase whose transport properties are consistent with those of an electronic nematic. We show that this phase, which exists at low temperatures in a narrow range of magnetic fields, forms directly from a quantum critical state, and contains more entropy than mean-field calculations predict. Our results suggest that this extra entropy is due to remnant degrees of freedom from the highly entropic state above $T_c$. The associated quantum critical point, which is 'concealed' by the nematic phase, separates two Fermi liquids, neither of which has an identifiable spontaneously broken symmetry, but which likely differ in the topology of their Fermi surfaces.*

# Introduction

One of the most striking empirical facts about quantum criticality is that, in systems with low disorder, the approach to QCPs is often cut off by the formation of new broken symmetry phases. Although this phase formation is widely discussed (1-4), thermodynamic data probing how it occurs are surprisingly sparse. The properties of a low temperature ordered phase are usually linked to those of the metal from which it condenses. Many states form from the background of well-understood Fermi liquids, so investigations of the metal are used to gain insight into the properties of the ordered phase. For example, the existence of sharply defined quasiparticles in a simple (Fermi liquid) metal implies the well known 'Cooper instability' that leads to the formation of a low temperature superconducting state, and the spectrum of phonons and/or magnetic excitations determines the structure of the gap function in that state. The case of phase formation from a quantum critical background is more challenging and possibly richer, since the metal itself is so mysterious; understanding the thermodynamics of phase formation might yield insight into the quantum critical metal as well as the ordered phase.

In-depth studies of the specific heat are difficult in prototypical quantum critical superconductors such as $CeIn_3$ and $CePd_2Si_2$ because of the need to work in pressure cells. Both constructing measurement apparatus suited to the high pressure environment and subtracting the huge addenda background due to the pressure cell are challenging experimental problems that have not yet fully been solved. In this paper we give a concrete example in which the material can be tuned through the quantum critical regime using magnetic field rather than pressure as the tuning parameter, enabling comprehensive measurement of the specific heat.

Interest in careful experimental studies of quantum criticality in metals has been further stimulated by a recent theoretical development that cuts across fields of physics, the so called "AdS-CFT" correspondence, based on dualities between conformal field theories (which presumably describe quantum critical systems) and a higher dimensional quantum gravity. In some cases, the quantum gravity is solvable, even though the conformal field theory is strongly interacting and hence impossible to analyze directly (5). As a result, novel strongly interacting "non-Fermi-liquid" critical theories have been characterized, and correspondingly the range of possible critical behaviours that can be imagined has been expanded. It is presently unclear whether any of the new critical theories apply to any realizable condensed matter system. Thermodynamic data from quantum critical materials are particularly desirable in order to investigate this possibility, and to place experimental constraints on the developing theories.

Here, we report the results of a study of the heat capacity of the field-tuned quantum critical system $Sr_3Ru_2O_7$, the *n*=2 member of the $Sr_{n+1}Ru_nO_{3n+1}$ Ruddlesden-Popper series of layered ruthenates. It consists of strongly coupled Ru-O bilayers, weakly coupled together to form a material with quasi two-dimensional conduction, based on hybridised Ru 4*d* – O 2*p* bands (6). The purity of our crystals can be characterized in terms of the residual ($T \rightarrow 0$) resistivity at *H*=0, which we express in terms of an inferred mean-free path, $\ell$. Samples with $\ell$ ~ 300 Å display transport properties consistent with the existence of a field-tuned QCP at $\mu_oH_c$ = 7.9 tesla and ambient pressure (7,8). In still purer samples with an order of magnitude larger mean-free path, the approach to this QCP is cut off by the formation of a new phase (9,10) associated with anisotropic magneto-transport suggesting the presence of nematic electronic order (11,12). A schematic summary of the main features of $Sr_3Ru_2O_7$ established by previous work is shown in Fig. 1. Although the QCP is hidden by the new phase, its fluctuations still dominate the broader phase diagram. $Sr_3Ru_2O_7$ is therefore an ideal system in which to study the thermodynamics of phase formation against a background of quantum criticality.

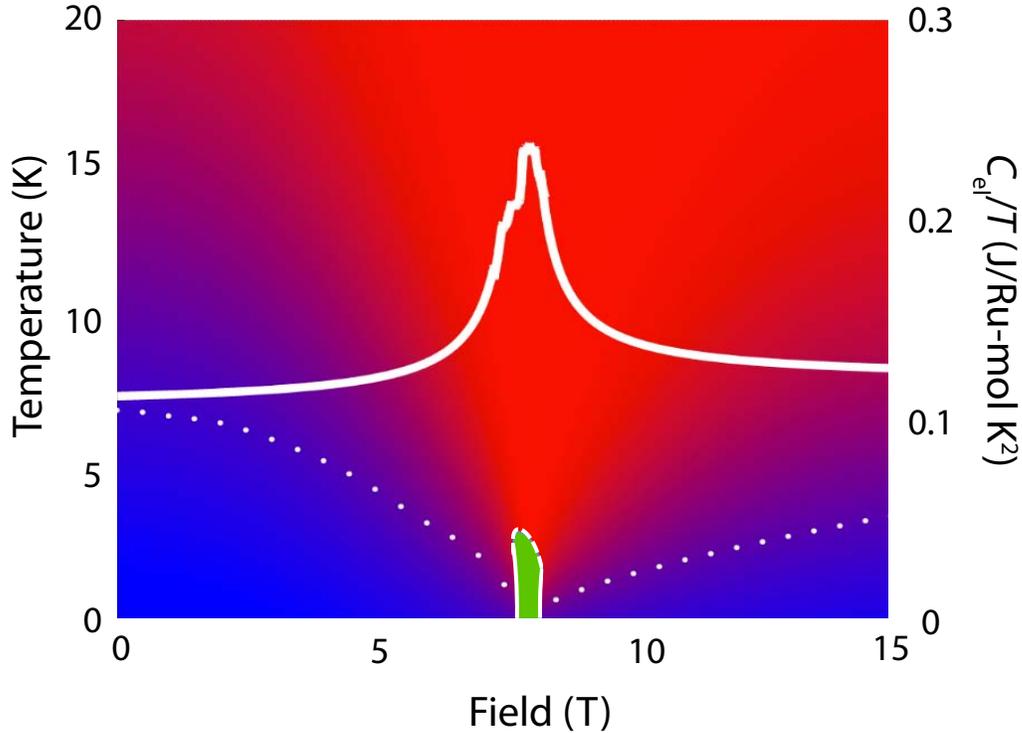

**Figure 1:** A schematic phase diagram for $Sr_3Ru_2O_7$ with magnetic field applied parallel to the crystallographic *c* axis, based on a combination of transport (7,9,10,11,13), thermal expansion (14), nuclear magnetic resonance (15) and quantum oscillations (16). Below a crossover temperature $T^*$ sketched by the dotted white line, Fermi liquids are seen at both low and high magnetic fields (blue shading). $T^*$, which is defined by thermodynamic measurements (14, Fig. 3C below) is depressed towards $T=0$ at a critical field $H_c$ of approximately 7.9 tesla, accompanied by the appearance of non-Fermi liquid temperature dependence of transport and thermodynamic properties (red shading) and features in magnetisation (7,13). In zero applied field, $T^* \sim 10$ K, and the material has a substantial specific heat coefficient of 110 mJ/molRuK$^2$, corresponding to large quasiparticle band renormalisations of a factor 10-30 compared with the predictions of LDA band calculations (17,18). The solid white lines sketch the field dependence of the electronic specific heat at 250 mK. The specific heat coefficient rises sharply as $H_c$ is approached from both the low and high-field sides (19). All of these observations are consistent with the existence of a quantum critical point (QCP) at 7.9 T, but in the highest purity samples, with mean free paths of several thousand Å, the approach to the QCP is hidden by the formation of a new phase, which is associated with the onset of anisotropic transport (10,11). This is indicated by the green shaded region, entered by first order phase transitions at low temperatures (solid white boundary) and a continuous transition at high temperatures (dashed white boundary). The temperature scale of this ordered phase has been multiplied by a factor of two for visual clarity. Approaching the quantum critical region at low temperatures from the low-field side, the effective mass determined from the specific heat has an apparent divergence, $m^* \sim [(H-H_c)/H_c]^{-1}$ as a function of increasing $H$, which is then cut-off near where the nematic phase occurs. The formation of the nematic phase is then accompanied by a small jump in entropy, followed by a drop on exit at the high field side (19).

## Results

In Fig. 2 we show the in-plane resistance ($\rho$) and electronic specific heat ($C_{el}/T$) for $Sr_3Ru_2O_7$ cooling from 18 K to 250 mK at 7.9 tesla in a single crystal with $\ell \sim 3000$ Å. Entry into the ordered phase at $T_c$ = 1.2 K is marked by a kink in $\rho$ and a step in $C_{el}/T$, but the data at higher temperatures are equally striking. For over a decade of temperature, $\rho$ is nearly perfectly linear in $T$, with $C_{el}/T$ varying as ln$T$ over the same range. Somewhat similar behavior has been observed in association with quantum criticality in a variety of other materials. In the present case these functional dependences are obeyed with high accuracy all the way down to $T_c$.

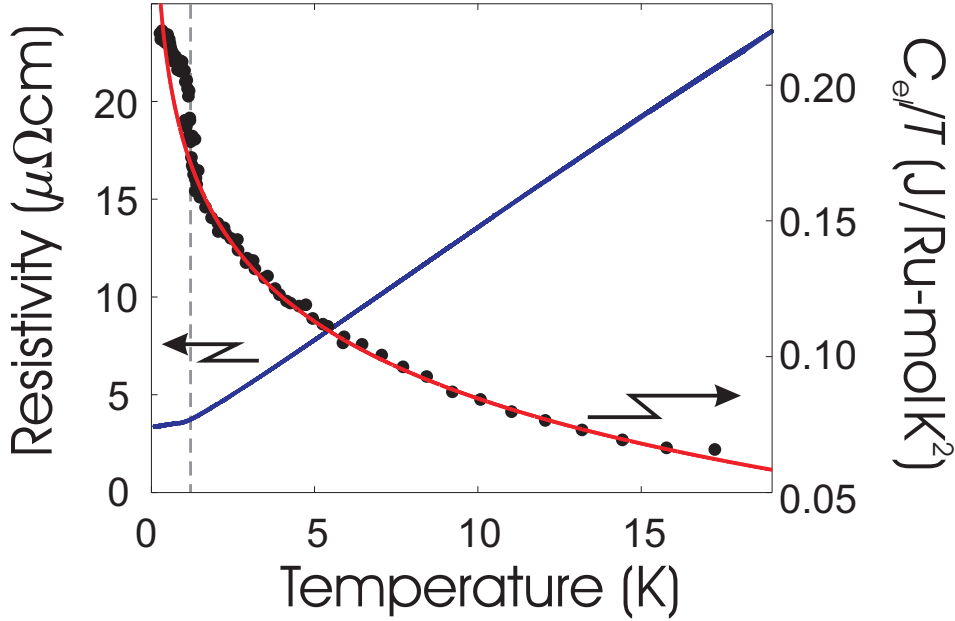

**Figure 2:** Resistivity (blue) and electronic specific heat (black dots) data for $Sr_3Ru_2O_7$ on cooling at the critical field of 7.9 tesla. The resistivity was measured between 100 mK and 18 K in a continuous run using an adiabatic demagnetisation refrigerator, and the results of both up and down sweeps are shown. The dotted grey line indicates the critical temperature of the nematic phase, and the red curve is a fit of the form $C_{el} = T\ln T$ to the data between 1.4 K and 18 K, extrapolated to 100 mK.

Before describing the phase formation in more depth, we turn our attention to the thermodynamic signatures of the hidden quantum critical point. Fig. 3A shows the evolution of the total specific heat below 40 K, in both zero applied field and at 7.9 tesla. Above 20 K, $C/T$ is field-independent, and can be fitted to high accuracy with the sum of a Debye model for the phonon contribution and a Fermi liquid constant term of $C/T$ = 70 mJ/Ru-mol K$^2$. Below 20 K, additional heat capacity is seen at both fields. The logarithmically diverging term highlighted in Fig. 2 dominates the data at 7.9 tesla, while a broad hump appears in zero applied field. This suggests that the quantum critical states might be formed from the depression towards $T$ = 0 of a low energy scale, $k_BT^*$, identified with the hump seen in zero field. As shown in Fig. 3B, this is what occurs.

As the field is increased, the hump sharpens and $T^*$ moves towards $T$ = 0 as $H$ approaches $H_c$, before beginning to grow again for fields higher than $H_c$. Plotting the position of the maximum for each of the fields studied demonstrates this depression and re-emergence of $T^*$ (Fig. 3C).

Further indications that the depression of a single energy scale is at the root of the quantum critical behaviour come from plotting the temperature dependence of the entropy at fields across the quantum critical region (Fig. 3D – details are given in the Supporting Information SI1). By 15K, just over 10% of $R$ln2 entropy per Ru is recovered at each of the applied fields*. Interpretations of this behaviour will be discussed below, but independent of these, the data in Figs. 2 and 3 provide compelling evidence that the $H$-$T$ phase diagram of Sr$_3$Ru$_2$O$_7$ is determined by the physics of quantum criticality.

As discussed below, features similar to those reported here have been observed in other quantum critical systems, although in those cases they have typically been associated with Kondo lattice physics which is not directly applicable to Sr$_3$Ru$_2$O$_7$. The fact that this quantum critical behaviour is observed with such clarity in this $d$-electron metal is one of our main experimental findings.

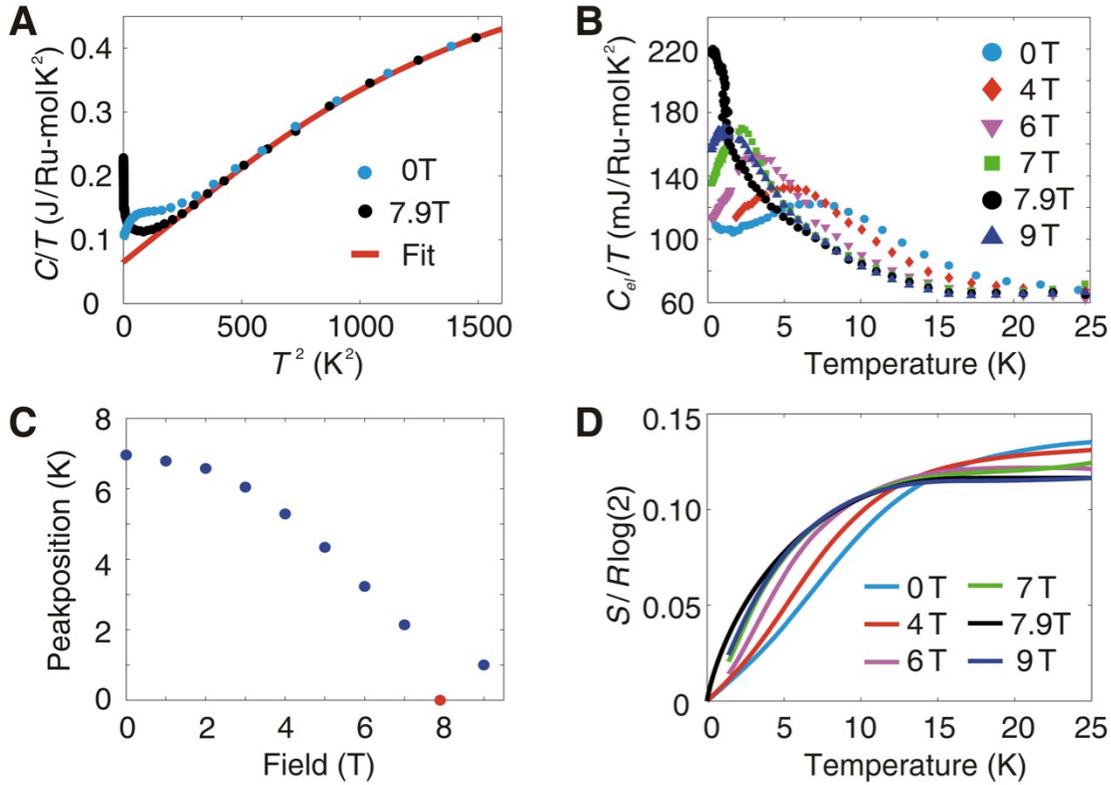

**Figure 3** A: Total specific heat divided by temperature for $Sr_3Ru_2O_7$ plotted against the square of temperature between 250 mK and 40 K in zero field (blue) and at 7.9 tesla (black). The red line is a fit to the field-independent data above 20K. It is the sum of a non-critical Fermi liquid component and a phonon contribution (16). B: The temperature dependence of the electronic specific heat (after subtraction of the phonon background) as the field is tuned through the quantum critical region. The hump seen in zero field sharpens and is depressed to progressively lower temperatures as the critical field is approached before reappearing on the high field side of the transition. The temperature at which the maximum occurs at each measured field is shown in panel N.B.: The y-axis values here are corrected for a typesetting error in the published version. C (blue dots). At 7.9 tesla no maximum is seen and based on previous data from more disordered samples it is assumed to occur at $T=0$ (red dot). In panel D we show the evolution of the entropy at each of the fields for which data are displayed in B.

The value of $T^*(H)$ inferred from the specific heat maximum correlates with the onset of a quadratic scattering rate in the resistivity for $T < T^*(H)$, as shown in the Supporting Information (SI2), as well as with a peak in the thermal expansion at $T \sim T^*(H)$ for $H$ close to $H_c$ (14), so it is reasonable to think of $T^*(H)$ as the scale of the renormalized Fermi temperature. By $H = H_c \pm 0.1$ tesla (the field width of the nematic phase) $T^*$ has fallen to below 1 K, so the transition into the phase at 1.2 K occurs not from a Fermi liquid but directly from a metallic state dominated by quantum critical fluctuations.

In Fig. 4A we show data for $C_{el}/T$ below 1.4 K at eight fields, four outside the ordered phase and four cooling down into it. Outside the phase, $C_{el}/T$ either falls slightly with decreasing $T$ (if the peak of Fig. 3B is at a low enough temperature that its falling edge still affects the data) or is approximately constant, as expected for a fully formed Fermi liquid. At the fields cooling into the phase, qualitatively different behaviour is seen. There is a step centred on approximately 1.2 K, followed by a $C_{el}/T$ that rises, approximately linearly, as the temperature drops further. The data are precise; as seen on the expanded scale of Fig. 4B, the midpoint of the step moves systematically down in temperature by approximately 60 mK as the field is increased from 7.9 tesla to 7.975 tesla. This shift is in excellent accord with the detailed shape of the phase diagram obtained previously using transport, thermal expansion and susceptibility (10). Here the most important point empirically is that the linear rise in $C_{el}/T$ is observed only within the ordered phase and never outside it. Linear terms in $C_{el}/T$ have been seen in strongly mass-renormalised Fermi liquids at low temperatures (20,21), and it is known that they can arise due to non-analytic terms in the Fermi liquid expansion (22). For that physics to be the source of our observations, however, the Fermi liquid parameters of the metals in and outside the ordered phase would have to differ strongly, something for which there appears to be little evidence (16).

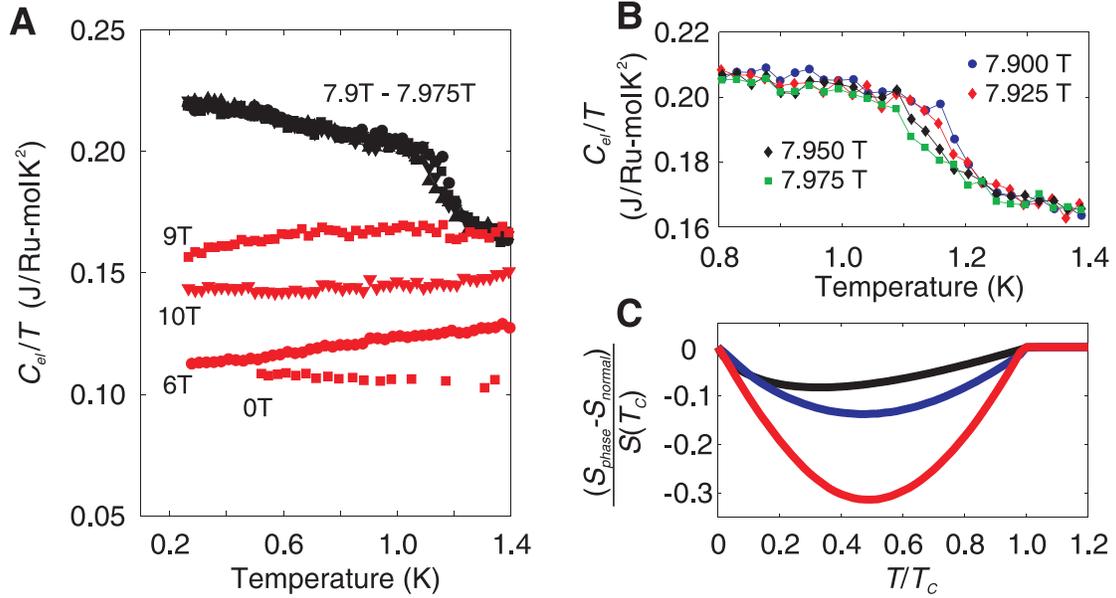

**Figure 4** A: Low temperature electronic specific heat divided by temperature for four fields above and below the nematic phase (red) and four within it (black). The transition into the phase is marked by the step-like feature centred on approximately 1.15 K. B: The transition region on an expanded scale, showing the systematic trend for the transition to be depressed to lower temperatures as the field is increased, in agreement with the known phase diagram (10). C: The entropy saving as a function of temperature at 7.9 tesla, expressed as a fraction of that at $T_c$ (black) compared with the results of two example model calculations: a phase opening a full gap (red) and a gapless Pomeranchuk distortion (blue). Full details of the models and the comparison can be found in the Supporting Information (SI4).

## Discussion

The data presented in Figs. 2 – 4 present a number of challenges to our understanding both of quantum criticality and of phase formation from a quantum critical background. The specific heat and entropy data of Figs 3B and 3D are similar to observations in classic heavy fermion compounds (2,3,23-27). In those *f*-electron systems, the local *f*-moments provide the entropy ($R\ln2$ per spin) of classical two-state fluctuators at high temperatures. As the temperature is lowered, these spins are incorporated into the Fermi sea via the Kondo effect, after which the entropy drops linearly with *T* at sufficiently low temperatures as implied by the Pauli principle restrictions of the Fermi liquid and the third law of thermodynamics. The energy scale associated with the hump in $C/T$ is associated with the Kondo temperature $T_K$ and the renormalized Fermi temperature $T_F$, the two being related. For reasons that are not fully understood, $T_K \rightarrow 0$ on the approach to heavy fermion quantum critical points, leading to a diverging effective mass (23,28). The tuning parameter for these heavy fermion QCPs can be pressure, chemical composition or magnetic field as in, for example, $YbRh_2Si_2$ and $CeAuSb_2$ (26,27). However, note that, in the majority of systems, the phase diagrams are 'asymmetric', with a magnetically ordered phase at low values of tuning parameter $h < h_c$ and a quantum disordered state for $h > h_c$.

In metamagnetic systems like $Sr_3Ru_2O_7$, the phase diagram is symmetric, with Fermi liquids for both $h < h_c$ and $h > h_c$, but we observe similar thermodynamics. A large entropy in excess of 10% of $R\ln 2$ per Ru begins to be quenched as the temperature drops below 20 K, with the drop becoming approximately linear at sufficiently low $T$ †. The closer the field is to 7.9 tesla, the larger the slope when the linear regime is reached, so the higher the effective mass of the resultant low temperature Fermi liquid. Previous field sweeps of $C_{el}/T$ and the magnetocaloric effect have directly established the existence of a Fermi liquid at 250 mK for fields below 7.5 tesla and above 8.5 tesla (19) with an appropriately field dependent effective mass, as summarised in Fig. 1; the current data show that this mass divergence is produced by the depression of the energy scale defined from the specific heat, $k_B T^*$.

Although the thermodynamic observations are similar to those made on heavy fermion systems, the microscopics cannot be, since $Sr_3Ru_2O_7$ has no $f$-electrons and hence no obvious source of fluctuating local moments, and no proximate ferromagnetic or antiferromagnetic phase on one side of the quantum critical point where these moments spontaneously order to form a broken symmetry state. Instead, we believe that a more general picture exists. We speculate that other, emergent low energy modes in the neighborhood of a quantum critical point can play qualitatively the same role. Above a characteristic energy scale which governs the quantum dynamics of the collective modes (and which vanishes at criticality), they are effectively classical and so make a contribution to the entropy which is large compared to that of the fermionic reservoir. However, at lower temperatures, the entire system forms a renormalized Fermi liquid, which therefore must have a heavy mass. The specific heat data of Fig. 2A suggest that a background fermionic reservoir and the low energy scale both exist in $Sr_3Ru_2O_7$. (Our specific heat data are also similar to observations in the $d$-band metal-insulator system $Fe_{1-x}Mn_xSi$ (29), although the disorder levels are much higher in that case.)

As illustrated in Fig. 4, phase formation in $Sr_3Ru_2O_7$ also has unusual properties. Some insight can be obtained into the observed behaviour by considering two different ways in which ordering affects the electronic structure of a system at mean-field level. One is by opening an energy gap on much or all of the Fermi surface. In this case, minimisation of $F = U - TS$ is achieved by making a large saving in $U$ that outweighs the cost of reducing the entropy below $T_c$. However, not all order involves gapping, and one would expect that in such situations, the minimisation of $F$ would be achieved without as strong a reduction of $S$ below $T_c$. As an illustration we have compared the measured temperature dependence of the entropy saving $\Delta S = S_{ord}-S_{norm}$ with the same quantity calculated in two simple mean-field models. (Here $S_{ord}$ is the measured entropy in the ordered phase, and $S_{norm}$ is the normal state entropy from above $T_c$ extrapolated into the ordered phase.) To represent the behavior of an ordered state which gaps the Fermi surface, we have computed $\Delta S$ in the mean-field description of the transition to an s-wave superconducting state produced by a weak, local attraction between electrons. As a model of a gapless ordered state, we considered the mean-field description of a Pomeranchuk transition to a nematic state with a distorted Fermi surface under circumstances in which the spin-up Fermi surface intersects the van Hove point. Indeed, in both models we have considered fine-tuned conditions in which a portion of the Fermi surface passes through a van Hove singularity so that the normal state $S/T$ is logarithmically divergent as T $\rightarrow$ 0, as it is in the experiment. (See Supporting Information (SI4) for the explicit details of the models). From the figure, one can see that in the fully gapped system (red), $\Delta S(T)$ has a deep minimum at $T=T_c/2$ where $\Delta S/S(T_c)$ = -0.31. By contrast, in the nematic model (blue), although the entropy still dips below $T_c$, the dip is much shallower, with a minimum of only $\Delta S/S(T_c)$ = -0.14 at $T\approx 0.5T_c$.

The similarity of the entropy data (black curve in Fig. 4C) to the model calculation for order without gapping strongly suggests that no gap opens at $T_c$ in $Sr_3Ru_2O_7$. However, the observed $\Delta S$ minimum is even shallower than that in the Pomeranchuk model. The real system has more entropy in the low temperature phase than this or other mean-field calculations predict (30-33). Although we cannot definitively rule out other sources for this entropy, the experiments reported here suggest that it is due to remnant degrees of freedom from the highly entropic state above $T_c$ ‡. The combination of these extra degrees of freedom and the logarithmically diverging normal state $C_{el}/T$ makes the low temperature entropy in the nematic phase higher than that of the adjoining Fermi liquids (19).

Combined with our previous work on the field dependence of the entropy at low temperatures, the results of this paper provide a comprehensive thermodynamic characterisation of a model quantum critical material supporting the formation of an unusual low temperature phase. A point worth emphasising is that our data do not appear to be compatible with a violation of the third law of thermodynamics such as that predicted in some AdS-CFT theories (35). Entropy is balanced within experimental error at $T_c$ if $C_{el}/T$ both above and below $T_c$ are extrapolated to $T$ = 0 (19). This implies that $S \to 0$ as $T \to 0$ within the nematic phase itself (Fig 3C). Moreover, the low temperature entropy of the hidden quantum critical point is associated with a relatively weak logarithmic divergence of $C_{el}/T$ that would not lead to a third law violation, even if it were not "hidden" by the nematic phase.

Although this paper is primarily concerned with experiment, we close by outlining a possible theoretical framework for future calculation. Previous proposals (30-32,36,37) for the origin of the nematic phase in $Sr_3Ru_2O_7$ have involved band electrons close to a van Hove singularity (vHs). As the magnetic field is increased, Zeeman splitting brings one spin species closer to the van Hove point, where even weak residual interactions are sufficient to produce broken symmetry phases such as nematic order, enabling the vHS to be effectively avoided. The underlying quantum critical point is thus associated with a change in the Fermi surface topology, rather than with a quantum transition between a broken symmetry phase and a disordered phase. Such models provide a natural source for the energy scale $k_BT^*(H)$, which is approximately the distance in energy between the Fermi level and the vHs for this spin species. In the Supporting Information, we analyse a specific model based on this simple picture and show that it captures some gross features of the behaviour of $Sr_3Ru_2O_7$, while others are not compatible with a simple rigid band Zeeman shift (38,39). It seems likely that in the real material, the quantum critical phase diagram is determined by the proximity to van Hove singularities coupled with one or more extra interaction terms, including the one which drives the nematicity, and ultimately acts to destroy the Fermi liquid behaviour in this system. In parallel with this class of theory, we believe that by determining a complete set of entropic data for this quantum critical system, our work facilitates quantitative tests of even more exotic critical theories constructed within the framework of AdS-CFT.

In summary, we have reported experiments addressing a problem of significance to condensed matter physics and beyond, namely the formation of a quantum critical state in a clean itinerant system and the phenomenology of ordered phase formation directly from such a state. The data point to a mechanism for the formation of heavy Fermi liquids that is more general than the spin-Kondo physics of conventional *f*-electron heavy fermion compounds, and give evidence for unconventional extra degrees of freedom in a phase formed from a quantum critical 'normal state'.

# Materials and Methods

Transport measurements were performed using standard four-terminal a.c. methods in a bespoke adiabatic demagnetisation cryostat that enabled continuous temperature sweeps from 100 mK to 20 K. Specific heat was studied above 600 mK in a commercial instrument (Quantum Design PPMS) equipped with a $^3$He cryostat, and between 250 mK and 1.4K in bespoke apparatus mounted on a dilution refrigerator.

In order to establish the field independent contribution to the specific heat $C$ we analysed the data above 20 K. We fitted it with a function of the form

$$\frac{C}{T} = \gamma + AT^2 \int_0^{T_D/T} \left\{ x^4 \frac{e^x}{(e^x - 1)^2} \right\} dx$$

with $\gamma$, $A$ and $T_D$ being free fit parameters and $T$ temperature. The first term ($\gamma$) represents a Fermi liquid contribution and the second term is a modified Debye model describing the phonon background. In order to allow for the more complicated phonon density of states of a realistic material we relaxed the condition that the prefactor $A$ is fixed by the Debye temperature $T_D$ and allowed them to vary independently.

**Footnotes**

\* Above this temperature there is increasing noise due to the range of integration combined with a weak field dependence consistent with the known temperature dependence of the magnetisation (see Supporting Information (SI3)).

† The fact that this entropy is 10% of $R$ln2 rather than the full value is not of crucial importance to the arguments given in the text, since not all states in the Brillouin zone need to be contributing to the specific heat peak.

‡ One possibility would be entropy associated with the domains that are thought to exist for this field orientation (30). However, estimates of the size of these domains makes them unlikely sources of as much extra entropy as we observe (16). A recent dilatometry study (34) appears to have settled the issue. Even in nearly monodomain samples in tilted field, the shape of the phase boundaries demonstrates that excess entropy remains in the nematic phase.

**Acknowledgements.** We acknowledge useful discussions with E. Fradkin, D.-H. Lee, C.A. Hooley and A.M. Berridge. APM thanks the Department of Physics at Stanford University for their hospitality during a sabbatical visit in which this manuscript was written, and the receipt of a Royal Society-Wolfson Research Merit Award. The work was supported by the UK EPSRC and Royal Society, and the US Department of Energy Grant # DE-FG02-06ER46287.

# Supporting Information

## SI1 – Calculation of Entropy based on Specific Heat and Magnetocaloric Effect

In principle the entropy *S* at temperature *T* and field *H* can be calculated from the specific heat *C* via the well-known relation

$$S(H,T) = \int_0^T \frac{C(H,T)}{T} dT \ .$$

(Eq. S1)

One important assumption in this calculation is that it is possible to make a reliable extrapolation of the measured specific heat to the experimentally inaccessible *T*=0 limit. This is for example feasible if the ground state of the material is a Fermi liquid and the experimental data extends well into the Fermi liquid regime such as in our case for magnetic fields up to 4 tesla. However, a general feature of quantum critical points is that the Fermi liquid regime itself is suppressed to below experimentally accessible temperatures in the vicinity of the quantum phase transition, arguably the most interesting part of the phase diagram. In the absence of a closed algebraic expression on which a controlled extrapolation of *C/T* could be based it is impossible to calculate the absolute entropy in the vicinity of a QCP using equation S1 alone.

For fields larger than 4 tesla we therefore carried out an analysis based on the magnetocaloric data reported previously (19). The detailed procedure is:

(1) Calculate the absolute specific heat at 4 tesla and 1.4K, *S*(4T, 1.4K), based on equation S1.
(2) Calculate at *T*=1.4K the isothermal difference between 4 tesla and the appropriate field, $\Delta S_{1.4K}(H)=S(H, 1.4K)-S(4T, 1.4K)$ based on data reported in (19).
(3) Use the following modification of equation S1 for the calculation of the entropy at any given field and temperature

$$S(H,T) = \int_{1.4K}^T \frac{C(H,T)}{T} dT \ + \Delta S_{1.4K}(H) \ + S(4T, 1.4K).$$

(Eq. S2)

# SI2 - Comparison of phase diagrams based on thermodynamic and transport data

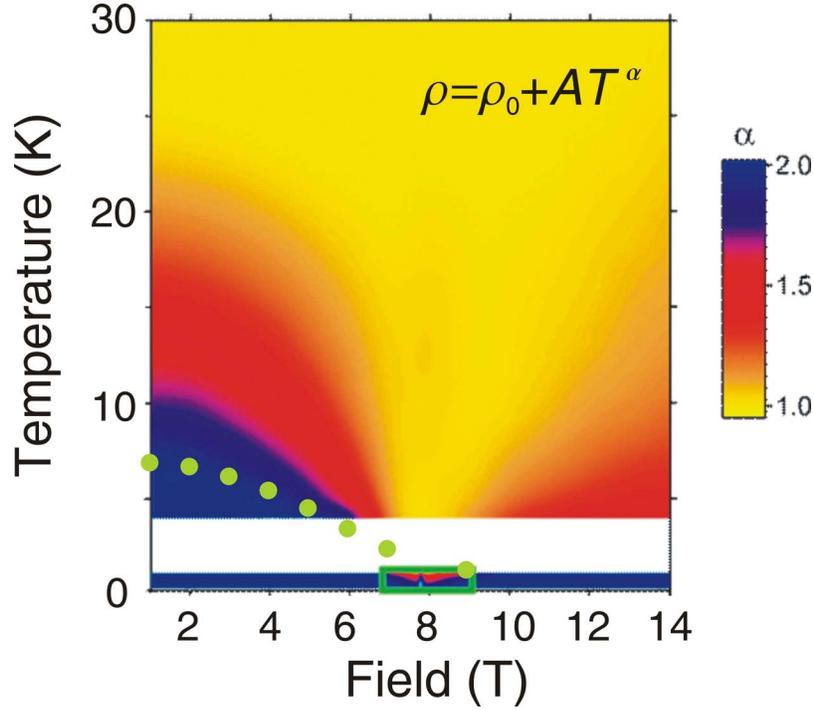

**Figure S1**: Comparison of the field-temperature phase diagram as extracted from resistivity and specific heat. The colour plot gives the exponent α of the expression $\rho = \rho_0 + AT^\alpha$ for the resistivity $\rho$ with $\rho_0$ being the elastic scattering contribution and $A$ a prefactor, both of which are temperature independent (partly reproduced from (7)). The data points (green) give the position of the maxima in $C_{el}/T$ as described in Fig. 3C. The crossover temperature $T^*$ to Fermi liquid behaviour in resistivity (α=2) has a similar field dependence to that extracted from the specific heat maximum (Fig. 3C main paper), and the two are also in fairly good quantitative agreement given that each is a somewhat arbitrary definition of $T^*$.

# SI3 - Estimate of field dependent entropy changes based on magnetic susceptibility

The small field dependence of the entropy seen in Fig. 3D of the main text is entirely consistent with the known behaviour of the magnetic susceptibility of $Sr_3Ru_2O_7$. As long as the susceptibility is field-independent, the magnetisation $M$ at temperature $T$ and magnetic field $\mu_0 H$ can be written as

$$M(T,H) = \int_0^H dH\, \chi(T,H) \approx \chi_{0T}(T) H$$

(Eq. S3)

The entropy $S$ can be expressed, via a Maxwell relation, as

$$\begin{aligned} S(T,H) &= \int_0^H dH\, \frac{\partial S}{\partial H} \\ &= \int_0^H dH\, \frac{\partial M}{\partial T} \end{aligned}$$

(Eq. S4)

Combining equations S3 and S4 gives

$$\begin{aligned} S(T,H) &= \int_0^H dH\, \frac{\partial}{\partial T}\{\chi_{0T}(T) H\} \\ &= \frac{1}{2} H^2 \frac{\partial \chi_{0T}}{\partial T} \end{aligned}$$

(Eq. S5)

At temperatures of 20 K and above, one can assume to first order that the magnetic susceptibility of $Sr_3Ru_2O_7$ is field independent up to magnetic fields of approximately 10 tesla. This can for example be seen in Fig. S2 (A) (reproduced from (40)). Based on the data shown in Fig S2 one finds that the temperature derivative of $\chi_{0T}$ at 20 K is approximately $1.7 \times 10^{-3}$ J/Ru-mol K T$^2$. This gives a contribution to the entropy at the critical field of $\approx 0.01\, R \log(2)$, consistent with the value obtained from the specific heat data and shown in Fig. 3D of the main paper. Inspection of Fig. S2 (B) further shows that the entropy would be expected to be roughly field-independent at 14K, as is seen in Fig. 3D (at this temperature significant field-dependent corrections to the susceptibility still exist, so the 'crossing point' would not be expected to be exact). At high temperatures, where field-linear susceptibility is expected, the thermal derivative of the magnetisation again becomes very small [Fig. S2 (B)], so the field dependence of the entropy will also be much reduced.

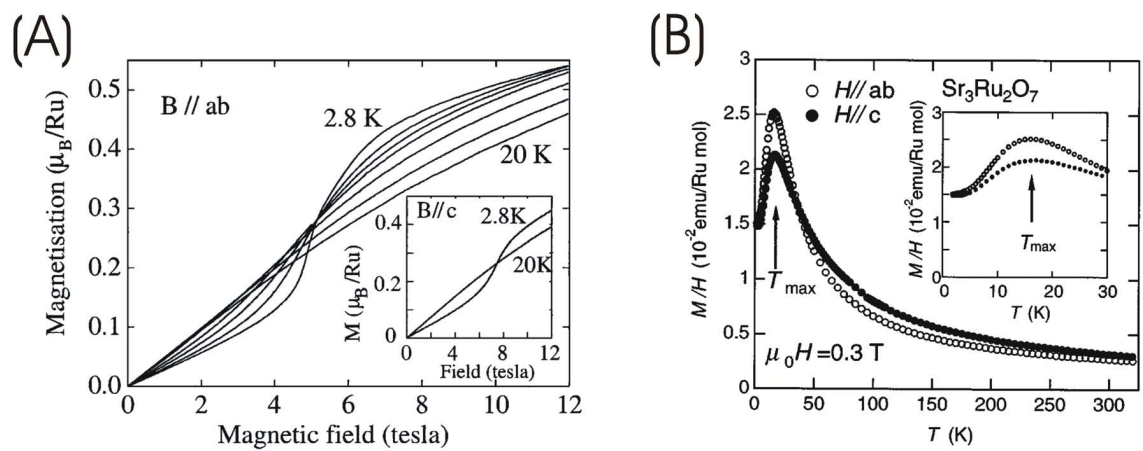

**Figure S2**: (A) Magnetisation as a function of magnetic field for several temperatures as indicated. Figure reproduced from (40). (B) Magnetic susceptibility as a function of temperature for different field orientations. Figure reproduced from (6).

# SI4 - Model calculation of nematic order and avoided preempted critical behaviour

*Introduction*

We have presented experimental data which suggests that a magnetic field induced quantum critical point (QCP) plays a crucial role in determining thermodynamic properties of $Sr_3Ru_2O_7$, even though the QCP itself is pre-empted by the formation of nematic order. The QCP is "exotic" in the sense that, as far as we know, it does not separate two phases that differ in symmetry. Here, we analyze an extremely simple model which exhibits, in caricature, the same sort of preempted quantum critical phenomena as seen in experiment. Specifically, we consider a model which exhibits a QCP in the limit that a certain interaction strength, *V*, is set equal to zero, but where the QCP is narrowly pre-empted by a nematic phase in the presence of a small, nonzero *V*. Moreover, in this model, the QCP is associated with a "Lifshitz" transition at which the Fermi-surface topology changes (*i.e.* the Fermi surface passes through a van Hove singularity), but no symmetry is broken; this is a deceptively simple example of an "exotic" quantum phase transition.

We will highlight not only those aspects of the experiment that are remarkably well captured by this simple model, but also those that are qualitatively different. These suggest that the ultimate quantum critical fixed point is fundamentally different than the one that we have analysed, but we argue below that analysis of this simplest case may provide a reasonable zeroeth order description of the underlying physics.

*Lifshitz Quantum Critical Point*

Following previous theoretical treatments of nematic order in this system, we start by considering tight-binding electrons on a square lattice in the presence of a Zeeman field.

$$\mathcal{H}_0 = \sum_{k,\sigma} \left[-2t(\cos k_x + \cos k_y) - 4t' \cos k_x \cos k_y - \sigma H - \mu\right] c^\dagger_{k,\sigma} c_{k,\sigma}$$

(Eq. S6)

where, $t, t'$ are respectively the nearest and next-nearest neighbor hopping integrals, and the chemical potential $\mu$ is tuned such that the Fermi surface lies close to the van Hove singularity (vHs). Our model closely follows that introduced by Yamase and Katanin (41, 42). A small non-zero $H$ moves one of the spin species closer to the vHs and the other spin species farther away from it. The density of states (dos) of each spin species, $\rho_\sigma$, has the following form for $H, \mu \ll \mu_c = 4t'$:

$$\rho_\sigma \sim \frac{1}{2\pi^2 t} \ln\left|\frac{t}{\mu - \sigma H - \mu_c}\right|.$$

(Eq. S7)

The QCP occurs when $\mu - \sigma H - \mu_c$, and the Fermi surface of one of the spin species crosses the vHs. Without loss of generality, we suppose that the "spin-up" Fermi sea ($\sigma = 1$) is tuned across the vHs.

This is a QCP in the precise technical sense that it corresponds to a point of non-analyticity,

$$E_0 \sim \frac{(\mu - \sigma H - \mu_c)^2}{4\pi^2 t} \ln\left|\frac{t}{\mu - \sigma H - \mu_c}\right|,$$

(Eq. S8)

of the ground-state energy (*i.e.* the $T \to 0$ limit of the free energy). In a more directly physical sense, it produces a well-defined quantum critical region in the phase diagram bounded by a field-dependent energy (temperature) scale $\omega^*(H) = |\mu - \sigma H - \mu_c|$ which collapses at the QCP. To investigate the way in which this energy scale plays a role in determining the finite temperature thermodynamic properties of the system in the vicinity of the QCP, we have computed the specific heat of the non-interacting system, shown in Figure S3. It can be seen from the figure that at temperatures $T \gg T^*$, $C/T$ is field-independent and varies logarithmically as a function of temperature. However, for temperatures $T < T^*(H)$ the logarithmic divergence of $C/T$ is cut off, producing a characteristic "hump" reminiscent of the experimental observations reported in the paper. Indeed, it can be seen that as the system approaches the QCP, $T^* \to 0$ and $C/T$ appears to diverge. By identifying the temperature $T^*$ at which the maximum in $C/T$ occurs, we are able to map out a phase diagram of the non-interacting system (Figure S5) where the crossover occurs from a logarithmic specific heat, reflecting the underlying vHs, to a constant $C/T$, reflecting the finite density of states of the system due to a non-zero $\omega^*$.

It is worth noting, as an aside, that naive scaling analysis of this QCP yields a value of the dynamical exponent, $z = 2$, from which it follows that spatial dimension $d = 2$ is the upper critical dimension. Correspondingly, there are apparently a number of interactions that are marginally relevant. One such interaction, on which we will focus, leads to a Pomeranchuk instability to a broken symmetry nematic phase. Others likely will give rise to non-trivial corrections to the quantum critical correlations or can lead to other broken symmetry phases. However, because these interactions are at most marginally relevant, it is plausible that the non-interacting QCP we analyze here can provide a reasonable zeroth order description of the physics. This is a point we intend to address in greater depth at a later date. We also draw attention to the weak second hump feature evident in Figure S3. This feature, which arises due to the contribution from the Zeeman split minority spin Fermi surface, is not present in the experimental data. Its absence calls into question the assumption of this and similar models that a rigid band Zeeman shift is a complete description of the observed physics in $Sr_3Ru_2O_7$.

***Effect of Weak Interactions***

The logarithmic divergence of the density of states allows for a wide range of electronic instabilities when one takes into account electron interaction effects. The interplay between electrons near a vHs and weak interactions has been studied extensively in the past by several authors (43,44), but continues to be an area with many unresolved questions. On a square lattice, there are many competing orders which can be stabilized, including density wave phases, orbital current phases, nematic phases, and even superconductivity to name just a few. We are not interested here in the phase diagram which results from a particular microscopic model near a vHs. Instead, we take a more phenomenological stance and focus on the instability towards the nematic phase alone, as is relevant in the case of $Sr_3Ru_2O_7$. Specifically, in addition to the non-interacting Hamiltonian discussed above, we add a set of short-ranged density-density interactions

$$\begin{aligned} \mathcal{H} &= \mathcal{H}_0 + \mathcal{H}_{int} \\ \mathcal{H}_{int} &= \sum_{i,j} V(\vec{r}-\vec{r}')\hat{n}_{\vec{r}}\hat{n}_{\vec{r}'} \end{aligned}$$

(Eq. S9)

where $\hat{n}_{\vec{r}} = \sum_\sigma c^\dagger_{\vec{r},\sigma} c_{\vec{r},\sigma}$ and $V(\vec{r}-\vec{r}') = V$ for nearest neighbor pairs and zero otherwise. We treat the interactions in the Hartree-Fock approximation and introduce a trial Hamiltonian with a nematic order parameter Δ:

$$\mathcal{H} = \mathcal{H}_0 + \Delta \sum_{k,\sigma} (\cos k_x - \cos k_y) c^\dagger_{k,\sigma} c_{k,\sigma}.$$

(Eq. S10)

The quantity Δ is a variational parameter that satisfies the following self-consistency relation:

$$\Delta = V \sum_{k,\sigma} (\cos k_x - \cos k_y) \langle c^\dagger_{k,\sigma} c_{k,\sigma} \rangle_{tr}$$

(Eq. S11)

where $\langle \hat{O} \rangle_{tr}$ denotes an expectation value computed in the trial ensemble. The self-consistency relations are solved at finite temperature and the resulting phase diagram is shown in Figure S5. There are several features of the phase diagram which we wish to emphasize. Firstly, the phase boundaries of the nematic phase are asymmetric with respect to the location of the preempted QCP, due to the broken particle-hole symmetry (i.e. $t' \neq 0$) in the system. Secondly, we note that for a range of magnetic fields surrounding the QCP, the nematic fluid phase forms from the "quantum critical" region where $C/T$ diverges logarithmically as the temperature is lowered; the divergence is cut off by the formation of the nematic phase. Lastly, note that in our mean-field calculation, the behavior of the slopes of the first order transitions into the nematic phase imply that the nematic phase has a lower entropy than the neighboring $C_4$ symmetric phases. This last point is inconsistent with the experimental observations, although to the best of our knowledge, it is a feature common to all Hartree-Fock descriptions of this transition.

A related issue is the behavior of $C/T$ at low temperatures inside the nematic phase. Figure S6 shows $C/T$ for the system for $V = 0.3t$ as a function of temperature at the critical field $H_C$. Significantly, it is seen that $C/T$ is an increasing function of temperature, in contrast with what is actually exhibited in our experiments. This is a crucial element of the experimental observations which is absent in our theoretical calculation.

*Summary*

In this Supporting Information, we have analyzed a simple model that surprisingly exhibits many features that are reminiscent of the experimental observations reported in this paper. We emphasize, however, that this model also fails to capture the striking deviations from Fermi liquid behavior exhibited by the material. Thus, we present here what we believe may be an adequate starting point for more sophisticated treatments, rather than any claim that the model that we analyze solves the problem.

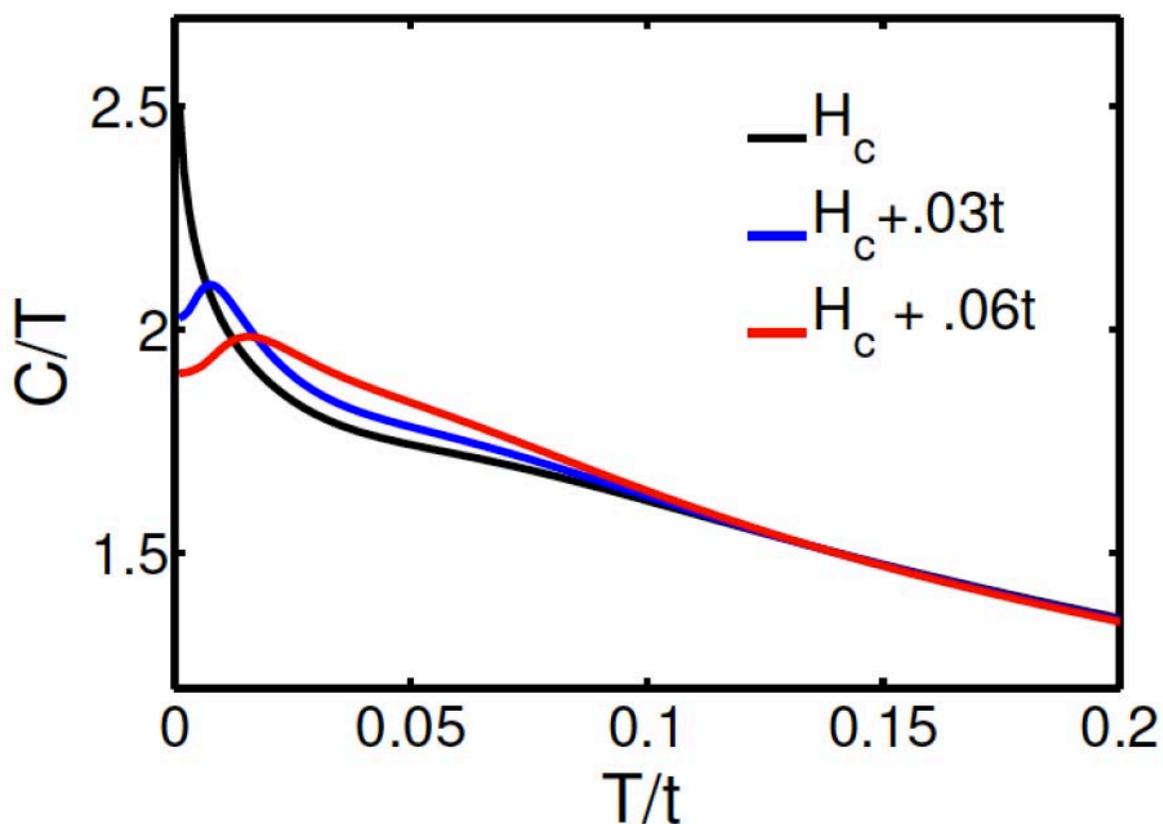

**Figure S3**: Specific heat of a non-interacting electron gas with $t = 1$; $t' = 0.3$; $\mu = 1.05$ at various magnetic fields. There is a hump in $C/T$ at a characteristic temperature $T^*$ above which $C/T$ varies logarithmically as a function of temperature. Note the second "shoulder" which is present in the curves which arises from the rigid band shift of the minority spin Fermi surface. The absence of this feature in the experimental data suggests that more is at play than a simple shift of weakly interacting electron states.

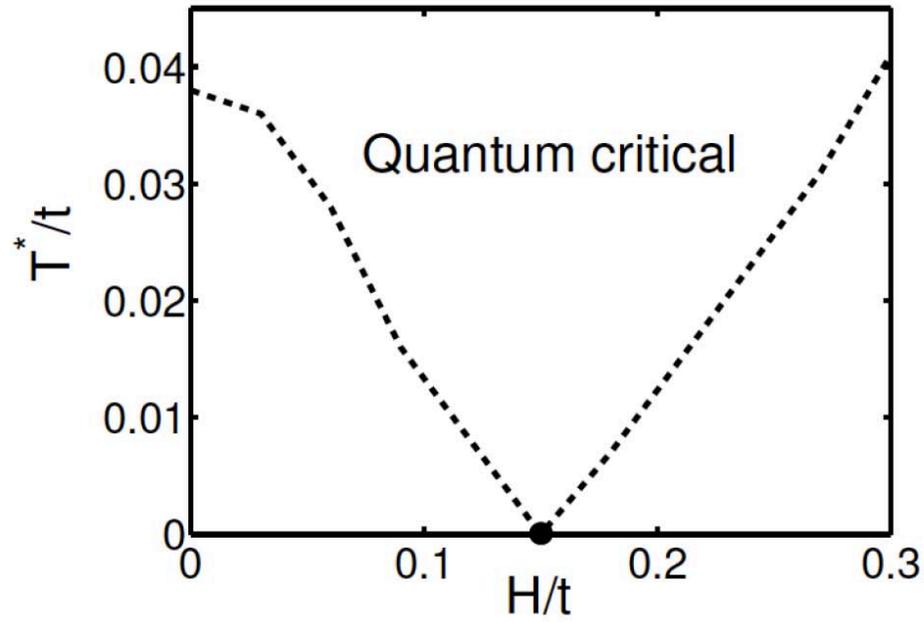

**Figure S4**: Behavior of $T^*$ as a function of field for the non-interacting system in Fig. S3.

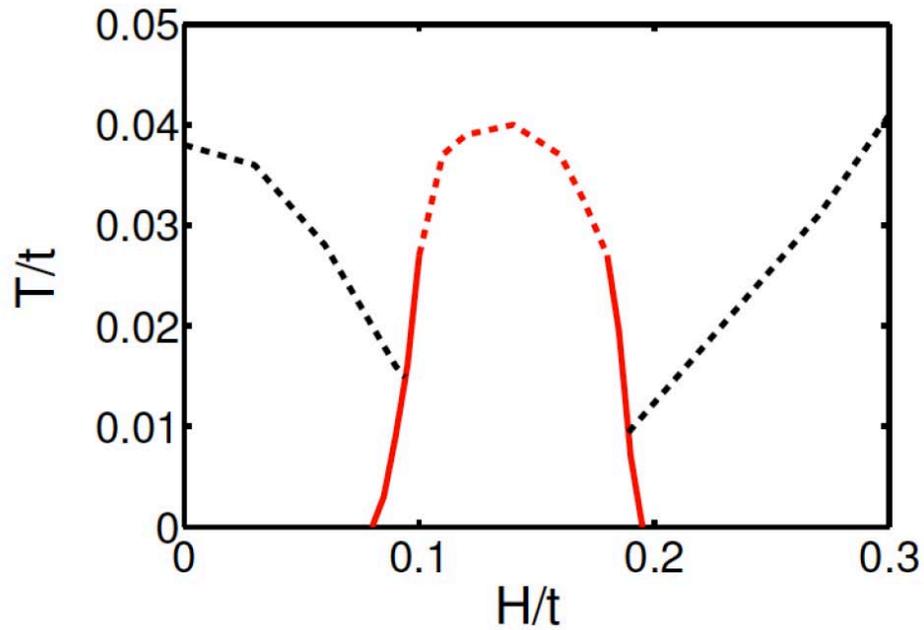

**Figure S5**: Phase diagram in the presence of weak-interactions, $V = 0.17t$. The quantum critical point is avoided by the formation of the nematic phase is shown in red. Also shown are the remaining crossovers arising from the avoided quantum critical point (dashed black curve). The solid red curve denotes first order transition whereas the dashed red curve denotes a continuous transition.

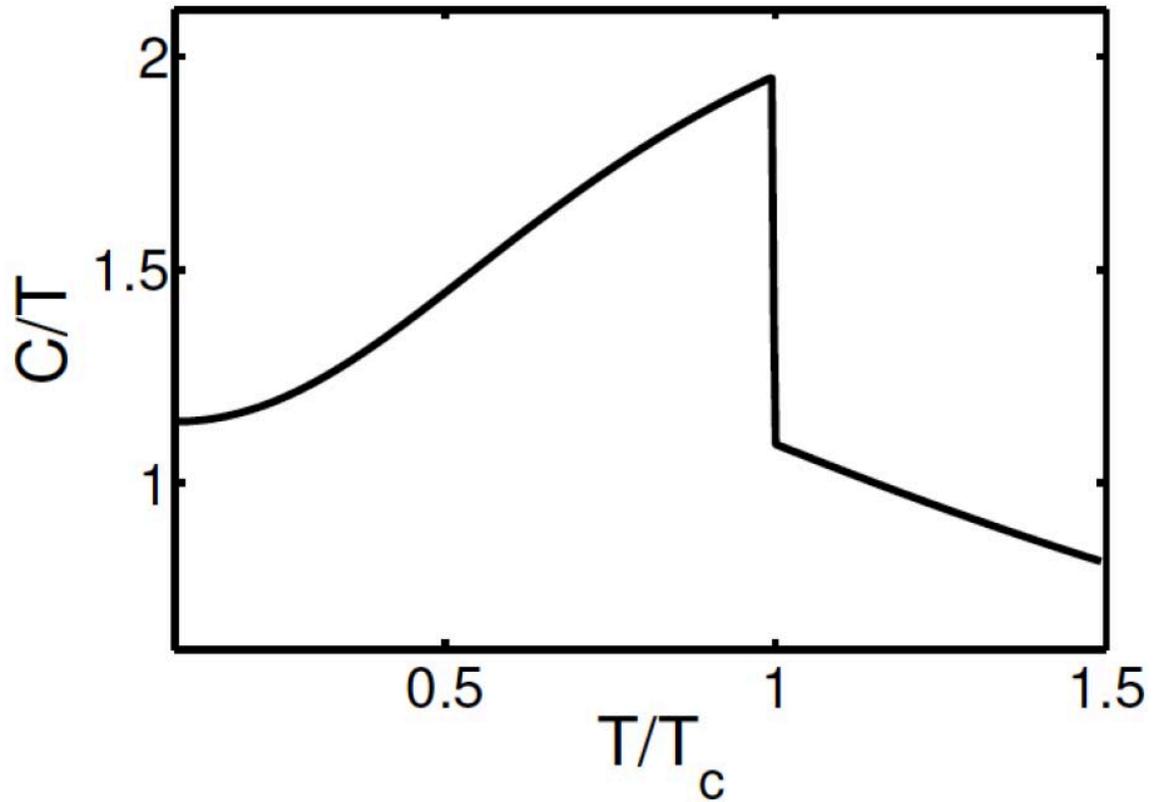

**Figure S6**: Specific heat $C/T$ as a function of temperature for $V = 0.3t$; $\mu = 1.05t$; $H = H_C = 0.15t$. For these parameters, the low temperature phase is a nematic phase which is destroyed at a temperature $T_C$. Note that $C/T$ is an increasing function of temperature, in contrast to what is found in our experimental observations.

**References**

1. Mathur, ND, et al. (1998) Magnetically mediated superconductivity in heavy fermion compounds. *Nature* 442: 39-43.

2. Löhneysen Hv, Rosch A, Vojta M, Wölfle P (2007) Fermi-liquid instabilities at magnetic quantum phase transitions. *Rev. Mod. Phys.* 79:1015-1075.

3. Stewart GR (2001) Non-Fermi-liquid behavior in d- and f-electron metals. *Rev. Mod. Phys.* 73:797-855.

4. Yang Y-F, Fisk Z, Lee H-O, Thompson JD, Pines D (2008) Scaling the Kondo lattice. *Nature* 454: 611-613.

5. For a review see: Sachdev S (2010) Condensed matter and AdS/CFT, http://arxiv.org/abs/1002.2947.

6. Ikeda S, Maeno Y, Nakatsuji S, Kosaka M, Uwatoko Y (2000) Ground state in $Sr_3Ru_2O_7$: Fermi liquid close to a ferromagnetic instability. *Phys. Rev. B* 62:R6089.

7. Grigera SA, et al. (2001) Magnetic field-tuned quantum criticality in the metallic ruthenate $Sr_3Ru_2O_7$. *Science* 294:329.

8. Grigera SA, et. al. (2003) Angular dependence of the magnetic susceptibility in the itinerant metamagnet $Sr_3Ru_2O_7$. *Phys. Rev. B* 67:214427.

9. Perry RS, et al. (2004) Multiple first-order metamagnetic transitions and quantum oscillations in ultrapure $Sr_3Ru_2O_7$. *Phys. Rev. Lett.* 92:166602.

10. Grigera SA, et al. (2004) Disorder-sensitive phase formation linked to metamagnetic quantum criticality. *Science* 306:1155.

11. Borzi RA, et al. (2007) Formation of a nematic fluid at high fields in $Sr_3Ru_2O_7$. *Science* 315:214-217.

12. Fradkin E, Kivelson SA, Lawler MJ, Eisenstein JP, Mackenzie AP (2010) Nematic Fermi fluids in Condensed Matter Physics. *Annual Reviews of Condensed Matter Physics* 1:153.

13. Perry RS, et al. (2001) Metamagnetism and critical fluctuations in high quality single crystals of the bilayer ruthenate $Sr_3Ru_2O_7$ *Phys. Rev. Lett.* 86:2661-2664.

14. Gegenwart P, Weickert F, Garst M, Perry RS, Maeno Y (2006) Metamagnetic Quantum Criticality in $Sr_3Ru_2O_7$ studied by thermal expansion. *Phys. Rev. Lett.* 96:136402.

15. Kitagawa K, et al. (2005) Metamagnetic quantum criticality revealed by O-17-NMR in the itinerant metamagnet $Sr_3Ru_2O_7$. *Phys. Rev. Lett.* 95:127001

16. Mercure JF, et al. (2009) Quantum oscillations in the anomalous phase in $Sr_3Ru_2O_7$. *Phys. Rev. Lett.* 103:176401.

17. Tamai A, et al. (2008) Fermi surface and van Hove singularities in the itinerant metamagnet $Sr_3Ru_2O_7$. *Phys. Rev. Lett.* 101: 026407.

18. Lee J, et al. (2009) Heavy d-electron quasiparticle interference and real-space electronic structure of $Sr_3Ru_2O_7$. *Nature Physics* 5:800-804.

19. Rost AW, Perry RS, Mercure J-F, Mackenzie AP, Grigera SA (2009) Entropy landscape of phase formation associated with quantum criticality in $Sr_3Ru_2O_7$. *Science* 325:1360.

20. Heuser K, Scheidt EW, Schreiner T, Fisk Z, Stewart GR (2000) Low temperature specific heat of $CeRu_2Si_2$ at the field induced metamagnetic instability. *J. Low Temp. Phys.* 118:235.

21. Flouquet J, Haen P, Raymond S, Aoki D, Knebel G (2002) Itinerant metamagnetism of $CeRu_2Si_2$: bringing out the dead. Comparison with the new $Sr_3Ru_2O_7$ case. *Physica B* 319:251.

22. Betouras J, Efremov D and Chubukov AV (2005) Thermodynamics of a Fermi liquid in a magnetic field. *Phys. Rev. B* 72:115112 and references therein.

23. Aoki Y, et al. (1998) Thermal properties of metamagnetic transition in heavy-fermion systems. *J. Magn. Magn. Mater.* 177:271.

24. Gegenwart P, et al. (2010) Divergence of the Grüneisen parameter and magnetocaloric effect at heavy fermion quantum critical points. *J. Low. Temp. Phys.* 161:117.

25. Löhneysen Hv, Pfleiderer C, Pietrus T, Stockert O, Will B (2001) Pressure versus magnetic-field tuning of a magnetic quantum phase transition. *Phys. Rev. B* 63:134411.

26. Oeschler N, et al. Low-temperature specific heat of YbRh$_2$Si$_2$. *Physica B* 403:1254-1256.

27. Balicas L et al. (2005) Magnetic field-tuned quantum critical point in CeAuSb$_2$. *Phys. Rev. B* 72:064402.

28. Shishido H, Settai R, Harima H, Onuki Y (2005) A drastic change of the Fermi surface at a critical pressure in CeRhIn$_5$: dHvA study under pressure. *J. Phys. Soc. Jpn.* 74:1103.

29. Manyala N, DiTusa JF, Aeppli G, Ramirez AP (2008) Doping a semiconductor to create an unconventional metal *Nature* 454:976-980.

30. Raghu S, et al. (2009) Microscopic theory of the nematic phase in Sr$_3$Ru$_2$O$_7$. *Phys. Rev. B* 79:214402.

31. Lee W-C, Wu C (2010) Theory of unconventional metamagnetic electron states in orbital band systems. *Phys. Rev. B* 80:104438.

32. Yamase H, Jakubczyk P (2010) Singular nonordering susceptibility at a Pomeranchuk instability. *Phys. Rev. B* 82:155119.

33. Puetter CM, Rau JG, Kee H-Y (2010) Microscopic route to nematicity in Sr$_3$Ru$_2$O$_7$. *Phys. Rev. B* 81:081105.

34. Stingl C, Perry RS, Maeno Y and Gegenwart P (2011) Symmetry-breaking lattice distortion in Sr$_3$Ru$_2$O$_7$. Phys. Rev. Lett. 107:026404.

35. D'Hoker E and Kraus P (2010) Holographic metamagnetism, quantum criticality, and crossover behavior. *J. High Energy Phys.* 5:83 and references therein.

36. Kee HY, Kim YB (2005) Itinerant metamagnetism induced by electronic nematic order. *Phys. Rev. B* 71:184402.

37. Berridge AM (2011) The role of band structure in the thermodynamic properties of itinerant metamagnets. *Phys. Rev. B* 83:235127.

38. Iwaya K, et al. (2007) Local tunneling spectroscopy across a metamagnetic critical point in the bilayer ruthenate Sr$_3$Ru$_2$O$_7$. *Phys. Rev. Lett.* 99:057078.

39. Farrell J, et al. (2008) Effect of electron doping the metamagnet $Sr_{3-y}La_yRu_2O_7$. *Phys. Rev. B* 78:R180409.

40. Perry RS (2001) Metamagnetism and critical fluctuations in high quality single crystals of the bilayer ruthenate $Sr_3Ru_2O_7$. *Phys. Rev. Lett.*, 86:2661.

41. Yamase H, Katanin AA (2007) Van Hove singularity and spontaneous Fermi surface symmetry breaking in $Sr_3Ru_2O_7$. *J. Phys. Soc. Jpn.* 76:073706.

42. Yamase H, Katanin AA (2010) Addendum to "Van Hove singularity and spontaneous Fermi surface symmetry breaking in $Sr_3Ru_2O_7$. *J. Phys. Soc. Jpn.* 79:127001.

43. Schulz HJ (1987) Superconductivity and antiferromagnetism in the two-dimensional Hubbard model: scaling theory. *Europhys. Lett.* 4:609.

44. Dzyaloshinskii IE (1987) Superconducting transitions due to Van Hove singularities in the electron spectrum. *JETP* 66:848.